\documentclass[prl,aps,twocolumn,showpacs,floatfix,tightenlines]{revtex4}

\usepackage{amsmath}
\usepackage{amsfonts}
\usepackage{epsfig}
\usepackage[dvips]{color}
\usepackage{graphicx}
\usepackage{subfigure}

\begin{document}

\title{Quantum process tomography of a single solid state qubit}

\author{M.~Howard$^1$, J.~Twamley$^1$, C.~Wittmann$^2$, T.~Gaebel$^2$, F.~Jelezko$^2$, and J.~Wrachtrup$^2$}
\affiliation{$^1$National University of Ireland, Maynooth, Co. Kildare, Ireland\\$^2$University of Stuttgart, 3. Physical Institute, Stuttgart, Germany}
\date{\today}
\begin{abstract}
We present an example of quantum process tomography performed on a single solid state qubit. The qubit used is two energy levels of the triplet
state in the Nitrogen-Vacancy defect in Diamond. Quantum process tomography is applied to a qubit which has been allowed to decohere for three different time
periods. In each case the process is found in terms of the $\chi$ matrix representation and the affine map representation. The discrepancy
between experimentally estimated process and the closest physically valid process is noted.
\end{abstract}
\vskip 0.1cm
\pacs{03.67.-a, 61.48.+c, 73.21.-b}
\maketitle

Quantum Process Tomography (QPT) \cite{jmodopt,nielsenandchuang} is a method designed to experimentally determine complete information about a
quantum process (quantum channel). This knowledge can aid in identifying any sources of errors encountered in performing a quantum gate
\cite{nielsenandchuang}. QPT has been performed in nuclear magnetic resonance \cite{nielsenNMR}, in optical systems \cite{nambu,Altepeter}, and,
to the best of our knowledge, we here present the first such demonstration of QPT for an individual solid-state based qubit. Methods have been
developed to determine a process from a tomographically incomplete set of measurements \cite{incomplete}, as well as techniques to ascertain the
master equation describing time evolution of the system \cite{superoperators,lindblad}. There are two different methods with which one can
perform QPT. The first method is to examine the effect of the unknown operation on a \emph{complete basis} of input states. The second method is
to examine the effect of the channel on a larger state composed of the original system and an ancilla system \cite{Altepeter,jamiolkowski}. In
either case state tomography is used to ascertain the output state(s) and this information is used in conjunction with the known input state(s)
to derive the process. The former method will be applied in what follows.

The Nitrogen-Vacancy defect consists of a substitutional nitrogen impurity next to a vacancy in the diamond lattice. The resulting electron
configuration produces a spin triplet (S=1) state which can be manipulated by electron spin resonance and read out by optically detected
magnetic resonance (ODMR)\cite{thir,fift,sixt} (Fig. 1(a)). At low temperature, the spin longitudinal relaxation time, $T_1$, is on the order of
seconds and thus a single electron spin state can be detected \cite{thir}. Transverse relaxation times (or decoherence times), $T_2$, of up to
60 microseconds have been reported in literature \cite{twtw}, for samples with low nitrogen concentration. A controlled two-qubit quantum gate
in which the vacancy electron spin is hyperfine coupled to a nearby C${}^{13}$ nucleus, has recently been performed using this system
\cite{twon}.

Using quantum process tomography one can determine a completely positive and trace-preserving map $\mathcal{E}$ which represents the process
acting on an arbitrary input state $\rho$ :
\begin{equation}
\label{eq:chi} \mathcal{E}(\rho)=\sum_{m,n=0}^{d^2-1} \chi_{mn}\hspace{3pt} {\hat A}_m \rho {\hat A}^\dag_n,
\end{equation}
where the ${\hat A_m}$ are a chosen basis for operators acting on $\rho$, and $d=2^n$ for an n qubit system. The matrix $\chi$ completely
describes the process $\mathcal{E}$, and can be reconstructed from experimental quantum state tomographic measurements. To obtain $\chi$ it is
first necessary to apply the process to each member of a complete basis of input states (e.g $\rho_s=|\psi_s\rangle
\langle\psi_s|$,$|\psi_s\rangle=\{|0\rangle,|1\rangle,\frac{1}{\sqrt{2}} (|0\rangle+|1\rangle),\frac{1}{\sqrt{2}}(|0\rangle+i|1\rangle)\}$, for
a single qubit) and perform state tomography on the resulting output states. From these measurements $\lambda_{jk}$ can be specified
\cite{nielsenandchuang}, in the relation ${\cal E}(\rho_j)=\lambda_{jk}\rho_k$ where $\rho_j$ form a basis for density matrices. In order to
determine $\chi$ from $\lambda$ one operates on $\lambda$ with the pseudoinverse of $\beta$, where $\beta$ is derived theoretically from the
relation ${\hat A}_m \rho_j {\hat A}^\dag_n = \beta^{mn}_{jk} \rho_k$. Using this last relation and \eqref{eq:chi} we can see $\lambda_{ij}
=\sum_{mn} \beta^{mn}_{ij} \chi_{mn}$ and so inverting $\beta$ gives us $\chi$, as required.

The QPT experiment was performed at room temperature using diamond nanocrytsals obtained from type Ib synthetic diamond. Coherence times on the
order of a microsecond have been reported previously for this type of diamond. In order to perform many repetitions in a reasonable amount of
time, a sample with a relatively short coherence time was chosen. Diamond nanocrystals were spin coated on a glass substrate, and single
nanocrystals were observed with a homebuilt sample-scanning confocal microscope. In order to ensure the presence of a single defect in the laser
focus, the second-order coherence was measured using Hanbury-Brown and Twiss interferometer and the contrast of the antibunching depth was
determined. Microwaves were coupled to the sample by a ESR microresonator connected to a 40W travelling wave tube amplifier.

The basic energy level scheme of the molecule relevant to the experiment is shown in Fig. 1(a). The ground electronic triplet state consists of
the three spin sublevels corresponding to spin projections $m_S = \pm 1, 0$. Sublevels with  $m_S =0$ and $m_S =1$, were selected as qubit
states.
\begin{figure}[!ht]
    \begin{center}
    \subfigure{\includegraphics[height=32mm]{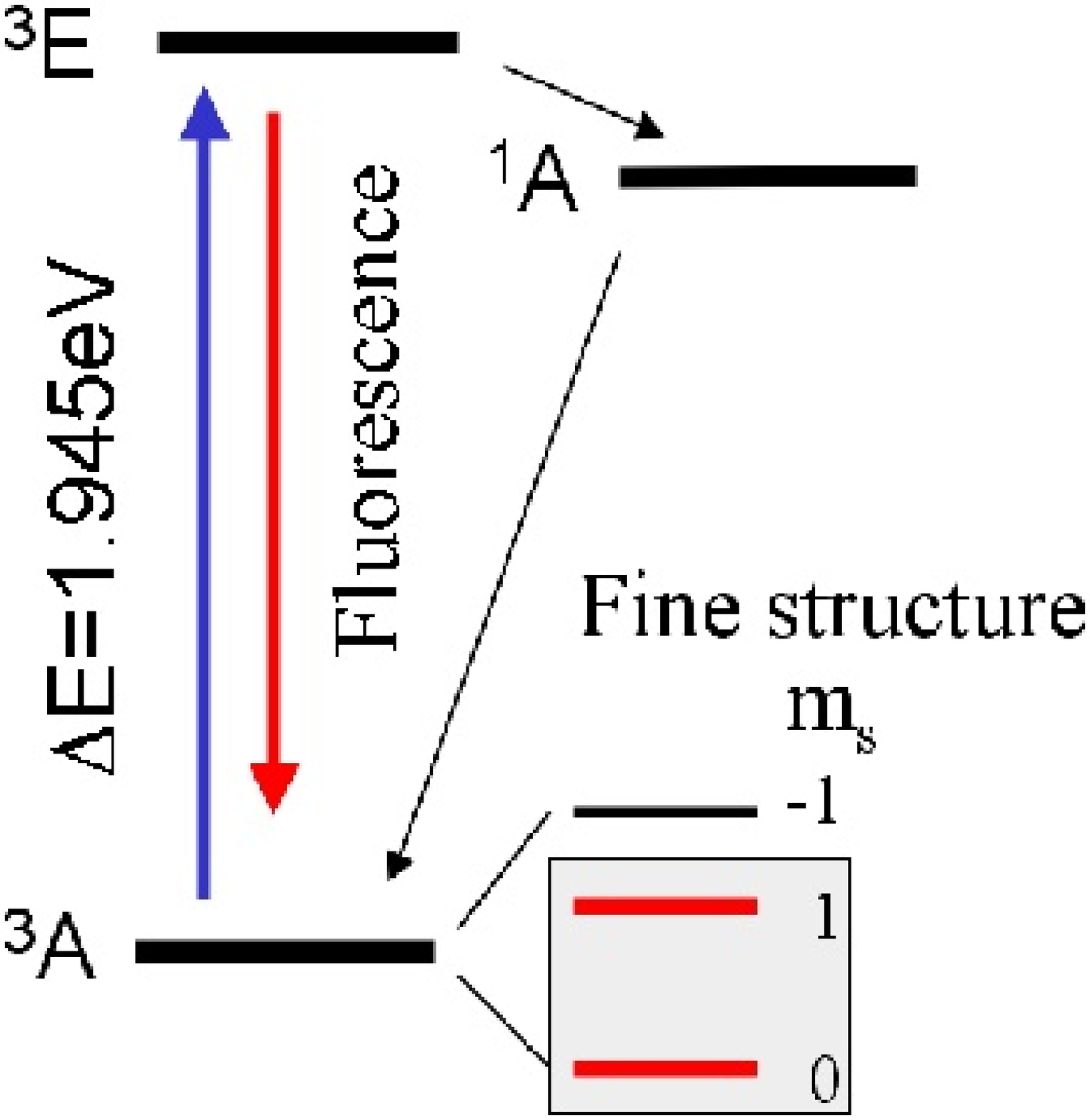}}{{(a)}}
    \vspace{10mm} \subfigure{\includegraphics[height=32mm]{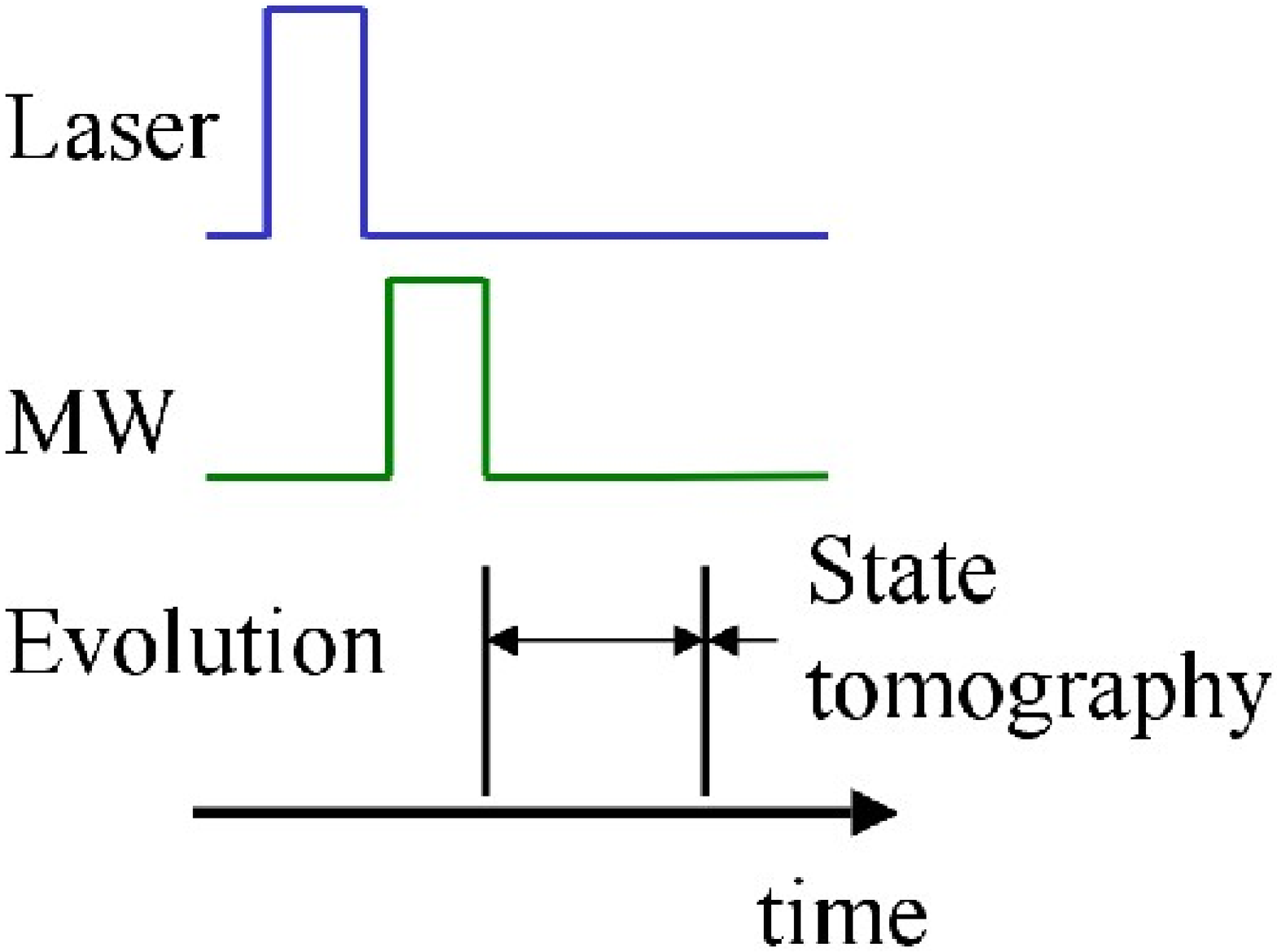}}{{(b)}}
  \caption{(a) Scheme of energy levels of NV center in weak magnetic field. Because of the C3v symmetry of the NV defect, the $m_S=\pm1$ sublevels remain degenerate in zero field and are split in non-zero magnetic field. (b) Experimental pulse sequence. Note that the laser was switched off during spin manipulation in order to avoid decoherence related to continuous spin measurement via optical excitation.}
\end{center}
\label{levels}
\end{figure}

In a preliminary step of the experiment, the times required to perform $\pi$ and $\frac{\pi}{2}$ pulses were determined using Rabi oscillations.
After that the experimental sequence shown in Fig. 1(b) was applied. As a first step the qubit was initialized in the $m_S=0$ state. This was
achieved by optical pumping, which results in a strong spin polarization, corresponding to at least 70\% population of the $m_S =0$ sublevel.
Optically induced polarization is related to spin-selective intersystem crossing from photoexcited ${}^3E$ triplet
 state to metastable singlet state ${}^1A$. A complete basis of four input states was then prepared using microwave pulses resonant
 with qubit transitions. The $\rho_{0}$ state  was obtained directly by optical pumping and three remaining input states $\rho_{1}$, $\rho_{2}$ and $\rho_{3}$
 were obtained by application of suitable $\pi$ or ${\frac{\pi}{2}}$ pulses. Each of these states was left to decohere for three different time intervals: 20 ns, 40 ns and 80 ns.
 As a last step, measurements of the diagonal and  off-diagonal elements of the density matrix were performed. Estimates of the density
matrix elements were extracted from experimental data (Rabi oscillations) using the maximum entropy (MaxEnt) technique \cite{MaxEnt}. This
method returns a physically valid density matrix which satisfies, as closely as possible, the expectation values of measured observables. In
cases where only incomplete knowledge of the output state is known, an additional constraint is used; the reconstructed state must also have the
maximum allowable von Neumann entropy.

The state of an arbitrary qubit can be expressed as $ \rho=\frac{1}{2}({\mathcal I} + \vec{r}\cdot\vec{\sigma})\leftrightarrow
\frac{1}{2}(1,x,y,z) $ where $\vec{\sigma}=\{\sigma_x,\sigma_y,\sigma_z\}$ and $x^2 + y^2 + z^2 \leq 1$. In this basis an arbitrary
trace-preserving process $\cal E$ can be written in the affine map form: $ {\cal E_{A}}=\left( \begin{array}{cc} 1&0 \\
\vec{t} & E
\end{array} \right)\ $
where $E$ is a real 3x3 matrix (responsible for deformation and rotation of the Bloch sphere) and $\vec{t}$ denotes displacement from
$\vec{r}=(0,0,0)$.

Using this picture of quantum processes \cite{incomplete,nielsenandchuang}, one can reconstruct the map ${\cal E}$ by examining the
transformations undergone by a complete basis of states: $\rho_0=\frac{1}{2}\mathcal{I},\ \rho_x =\frac{1}{2}(\mathcal{I}+\sigma_x),\ \rho_y
=\frac{1}{2}(\mathcal {I}+\sigma_y) $ and $ \rho_z =\frac{1}{2}(\mathcal{I}+\sigma_z)$. For brevity we write ${\cal
E_{A}}(\rho_0)=\vec{m'},{\cal E_{A}}(\rho_x)=\vec{x'},{\cal E_{A}}(\rho_y)=\vec{y'}$ and ${\cal E_{A}}(\rho_z)=\vec{z'}$.
The transformation ${\cal E}$ can then be expressed in the affine form by \begin{displaymath} {\cal E_{A}}=\left( \begin{array}{cccc} 1&0&0&0 \\
\vec{m'} &\vec{x'}-\vec{m'} &\vec{y'}-\vec{m'} &\vec{z'}-\vec{m'}
\end{array} \right)\
\end{displaymath}

A map is completely positive \cite{Choi,jamiolkowski} and trace preserving if and only if it can be written in the Kraus (operator sum)
representation: $\mathcal{E}(\rho)= \sum_{k} {\hat A}_{k} \rho{\hat A}^\dag_{k}$ where $\sum_k{\hat A}^\dag_k {\hat A}_k = \mathcal{I}$.
Physically, the complete positivity requirement states that, for a one-qubit process $\mathcal{E}$ and an entangled Bell state
$|\psi^+\rangle=\frac{1}{\sqrt{2}}(|00\rangle + |11\rangle)$, $(\mathcal{E} \otimes \mathcal{I}) |\psi^+\rangle$ must also be a valid two qubit
state.

The process obtained directly from experimental data is often unphysical i.e. non-trace-preserving or not completely positive. In such cases it
is necessary to search for a physical process which is closest in some sense to the measured experimental results. In this case we used a least
squares fit between the experimentally determined $\chi$ and a Hermitian parametrization \cite{dfvjQST} of a physical $\tilde{\chi}$ while
enforcing complete positivity and trace preservation.

The graphical representation of a process as a deformation and rotation of the bloch sphere (Fig. 2-4) can give crude but obvious insights into
the physicality or otherwise of the estimated experimental process (protrusions outside the unit bloch sphere as well as ellipsoids which are
``flattened" in one dimension betray a violation of the trace preservation and complete positivity conditions) and also into the fidelity of the
desired unitary process. It is important to note that not all ellipsoids within the bloch sphere can be obtained by a completely positive map.
Transverse relaxation (the dephasing channel) is depicted as the Bloch sphere collapsing to the z axis. As expected this is seen to be dominant
decoherence process. Longitudinal relaxation (amplitude damping) is not discernible on the timescales used in the experiment.

It is possible to convert the process $\mathcal{E}$ into a density matrix, $\rho_{\mathcal{E}}$, via the Jamiolkowski isomorphism
\cite{distance,jamiolkowski}: $\rho_{\mathcal{E}} \equiv \left[\mathcal{I} \otimes \mathcal{E} \right]\left(|\Phi\rangle \langle \Phi |\right)$
where $|\Phi\rangle= \sum_{j} |j\rangle |j\rangle / \sqrt{d}$ and $\{|j\rangle \}$ is an orthonormal basis. When the process $\mathcal{E}$ is
physical, one then obtains a physically valid $\rho_{\mathcal{E}}$ which can be compared to the ideal process (converted to $\rho_{id}$) using
distance measures on quantum states \cite{distance}. The \textbf{trace distance} between density matrices $\rho_{id}$ and $\rho_{\mathcal{E}}$
is $D(\rho_{id},\rho_{\mathcal{E}})\equiv \frac{1}{2} tr \mid \rho_{id}-\rho_{\mathcal{E}}\mid$, where $\mid X\mid \equiv \sqrt{X^\dag X}$.
 Similarly one can define the {\bf Fidelity}:
$F(\rho_{id},\rho_{\mathcal{E}})\equiv tr \left(\sqrt{\sqrt{\rho_{id}}\ \rho_{\mathcal{E}} \sqrt{\rho_{id}}}\right)^2$ and, using this
definition,
\begin{description}
 \item[Bures Metric] $B(\rho_{id},\rho_{\mathcal{E}})\equiv \sqrt{2-2\sqrt{F(\rho_{id},\rho_{\mathcal{E}})}}$ and

 \item[C metric] $C(\rho_{id},\rho_{\mathcal{E}}) \equiv \sqrt{1-{F(\rho_{id},\rho_{\mathcal{E}})}}$.
\end{description}
An unphysical process, however, can lead to an unphysical $\rho_{\mathcal{E}}$, possibly resulting in a process fidelity which is greater than
one. The application of the preceding fidelity-based distance measures can, therefore, produce nonsensical results. In such cases it is
necessary to use other techniques in order to estimate the disparity between $\chi$ and $\tilde{\chi}$ for example. If one defines $X
=\chi-\tilde{\chi}$, then possible measures are the matrix p-norms ($p=1,2,\infty$) of $X$ and the Frobenius norm of $X$ ($\|X\|_{Fro}$) as well
as the trace distance ($D_{pro}$).

As stated above, these quantities gives a measure of how well $\tilde{\chi}$ describes the experimental results. The fidelities corresponding to
these measures for the different decoherence durations are presented in Table I (note that $\|X\|_1 = \|X\|_{\infty}$ because $X$ is Hermitian
and that the basis operators used for $\chi$ (i.e. the operation elements) are $\left\{ \sigma_0, \sigma_x, -i\sigma_y, \sigma_z \right\}$).

\begin{table}
\begin{center}
\begin{tabular}{|c|c|c|c|c|}
  \hline
  Decoherence & & & & \\
  time&  $\|X\|_{p=1}$ & $\|X\|_{p=2}$ & $\|X\|_{Fro}$ & $D_{pro}$ \\
  \hline
  20ns &  0.101 & 0.050 & 0.066 & 0.056 \\
  40ns & 0.110 & 0.059 & 0.075 & 0.062 \\
  80ns & 0.175  & 0.075 & 0.110 & 0.096 \\
  \hline
\end{tabular}
\caption{The disparity between the estimated process from the experimental measurements ($\chi$) and the ``nearest'' physically valid process
  ($\tilde{\chi}$).}
  \end{center}\label{table1}
\end{table}

We have presented the first quantum process tomographic analysis of  an individual single solid-state qubit, a Nitrogen Vacancy centre in
Diamond.  This analysis is only possible due to the enormous advances made in recent years in single-molecule spectroscopy
\cite{thir,fift,sixt}, where the resultant ODMR technique provides us here with high-fidelity single-qubit readout. As quantum devices develop
and increase in size, the task of ``debugging" the device, or actively identifying the noise present in the device, will pose significant
challenges. The work presented here represents an initial step towards the testing of quantum devices in solid-state.

M.H. acknowledges funding from the Irish Research Council for Science, Engineering and Technology ``Embark" Initiative. The work was also
supported by the EC FP5 QIPC project QIPDDF-ROSES.
 \\

\begin{figure}[!ht]
    \begin{center}
    \subfigure{\epsfig{file=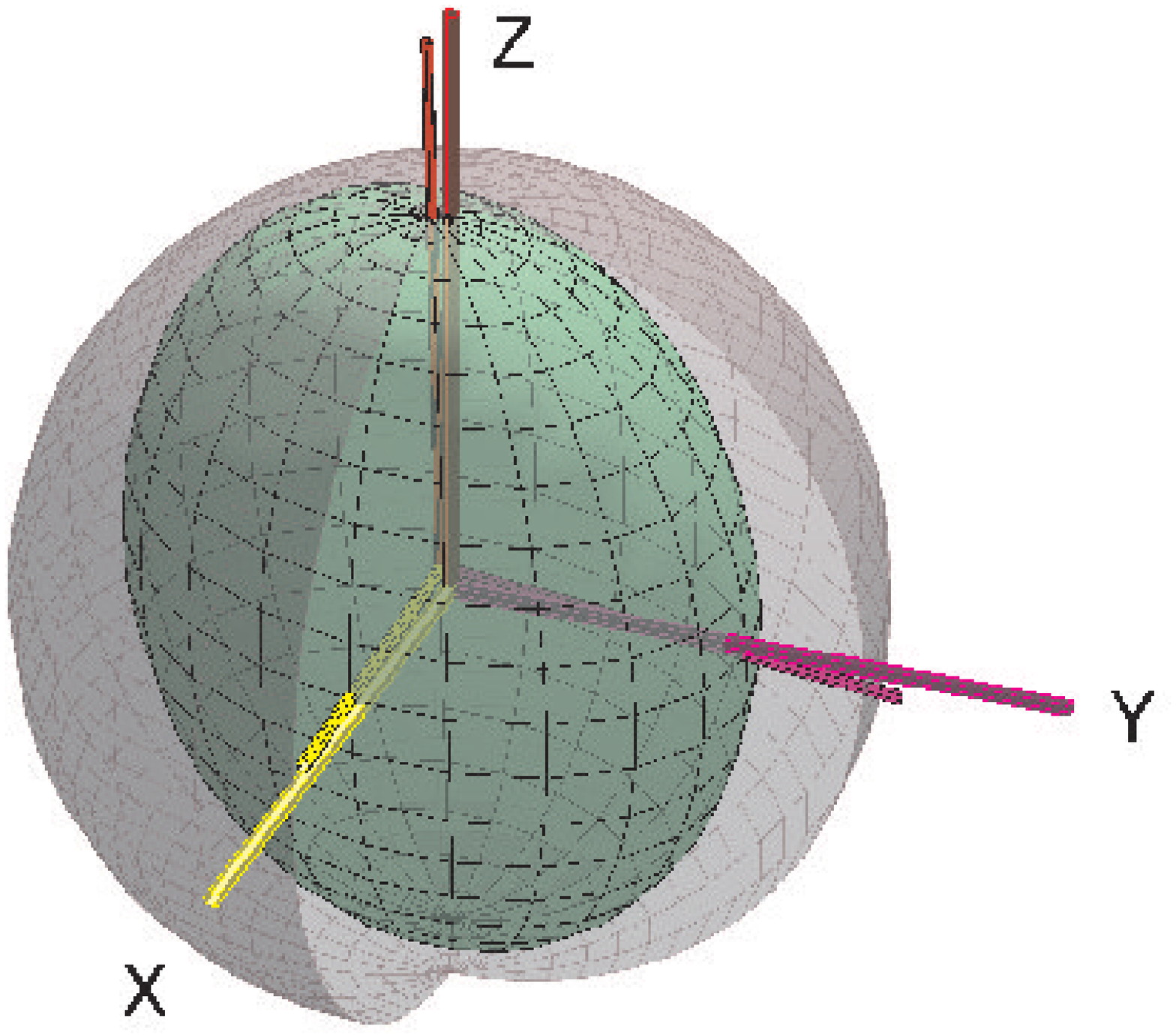,height=33mm}}{{(a)}}
    \subfigure{\epsfig{file=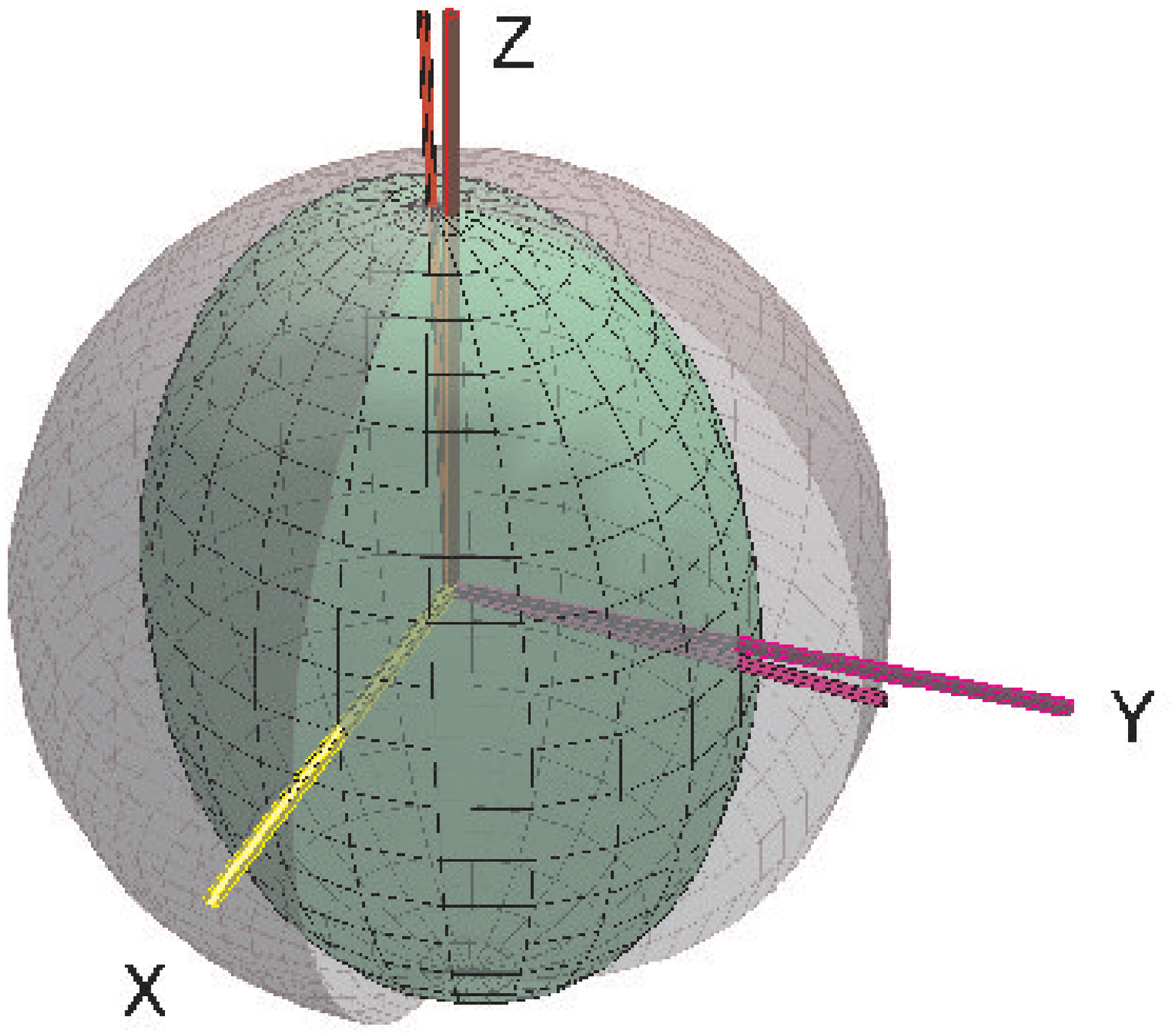,height=33mm}}{{(b)}}
\caption{20ns results:(a) Bloch sphere representation of process from experimental data. (b) Bloch sphere representation of process obtained
from $\tilde{\chi}$. }
\end{center}
\label{fig1}
\end{figure}

\begin{figure}[!ht]
    \begin{center}
    \subfigure{\epsfig{file=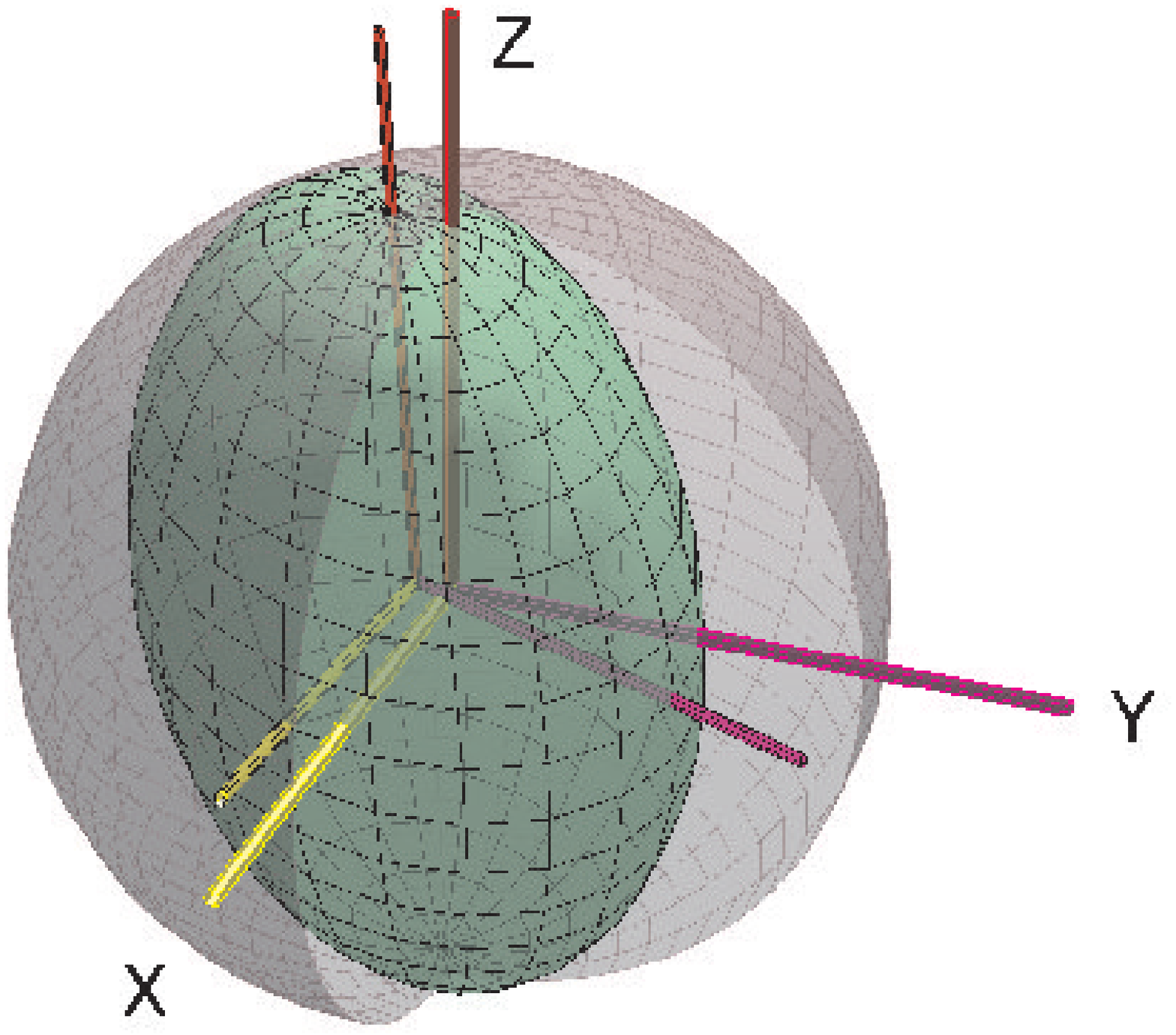,height=33mm}}{{(a)}}
    \subfigure{\epsfig{file=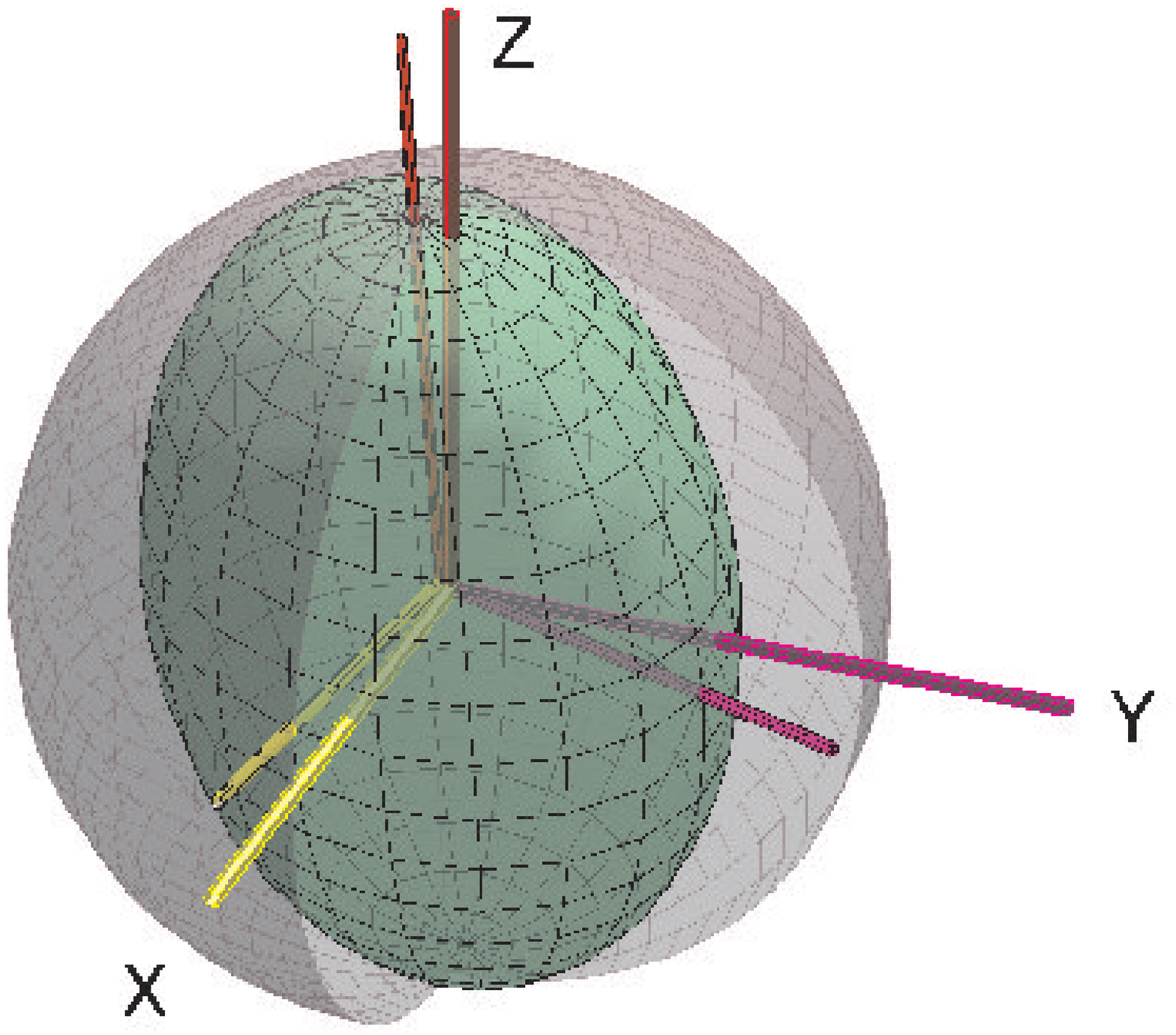,height=33mm}}{{(b)}}
\caption{40ns results:(a) Bloch sphere representation of process from experimental data. (b) Bloch sphere representation of process obtained
from $\tilde{\chi}$. }
\end{center}
\label{fig2}
\end{figure}

\begin{figure}[!ht]
    \begin{center}
    \subfigure{\epsfig{file=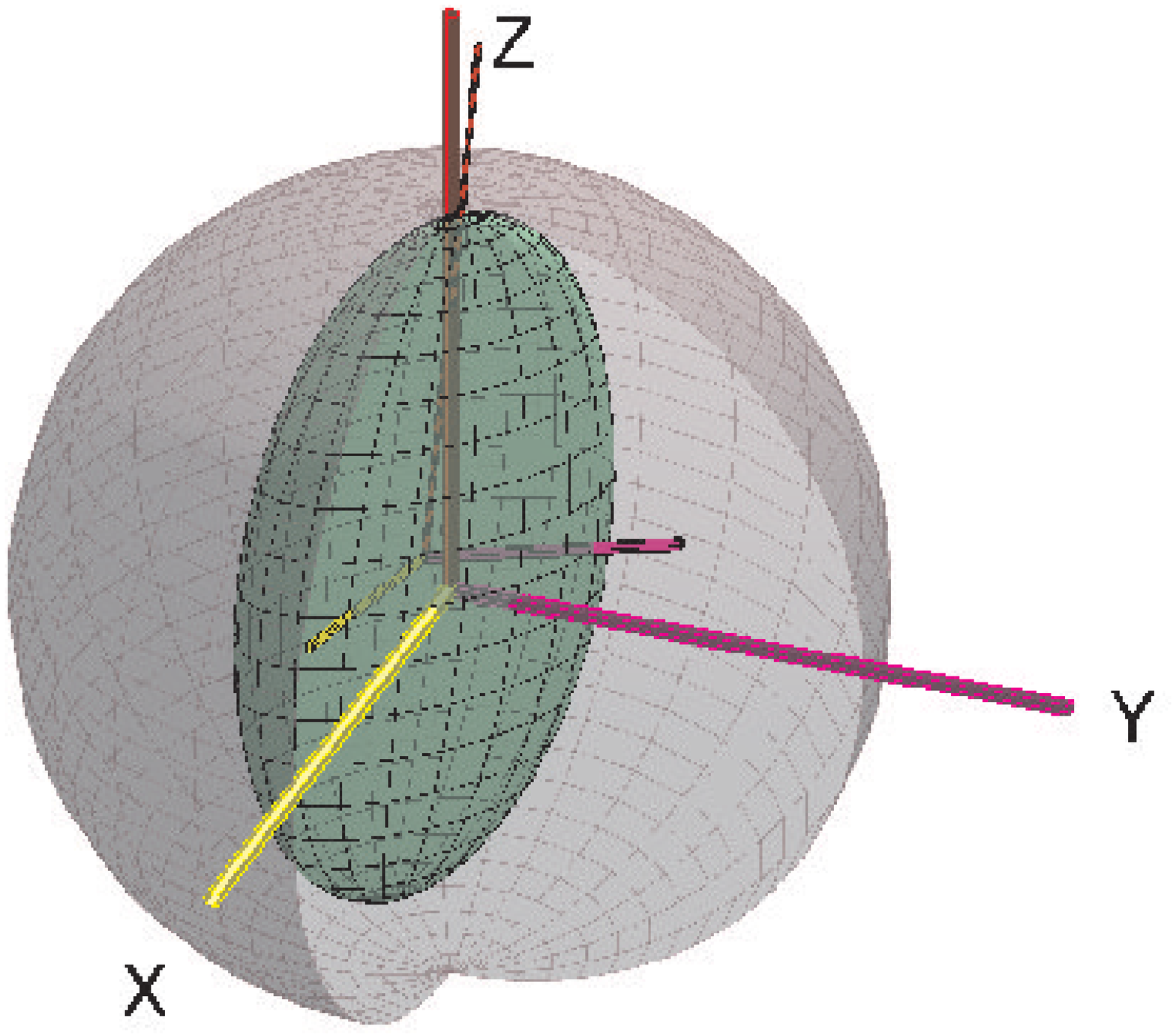,height=33mm}}{{(a)}}
   \subfigure{\epsfig{file=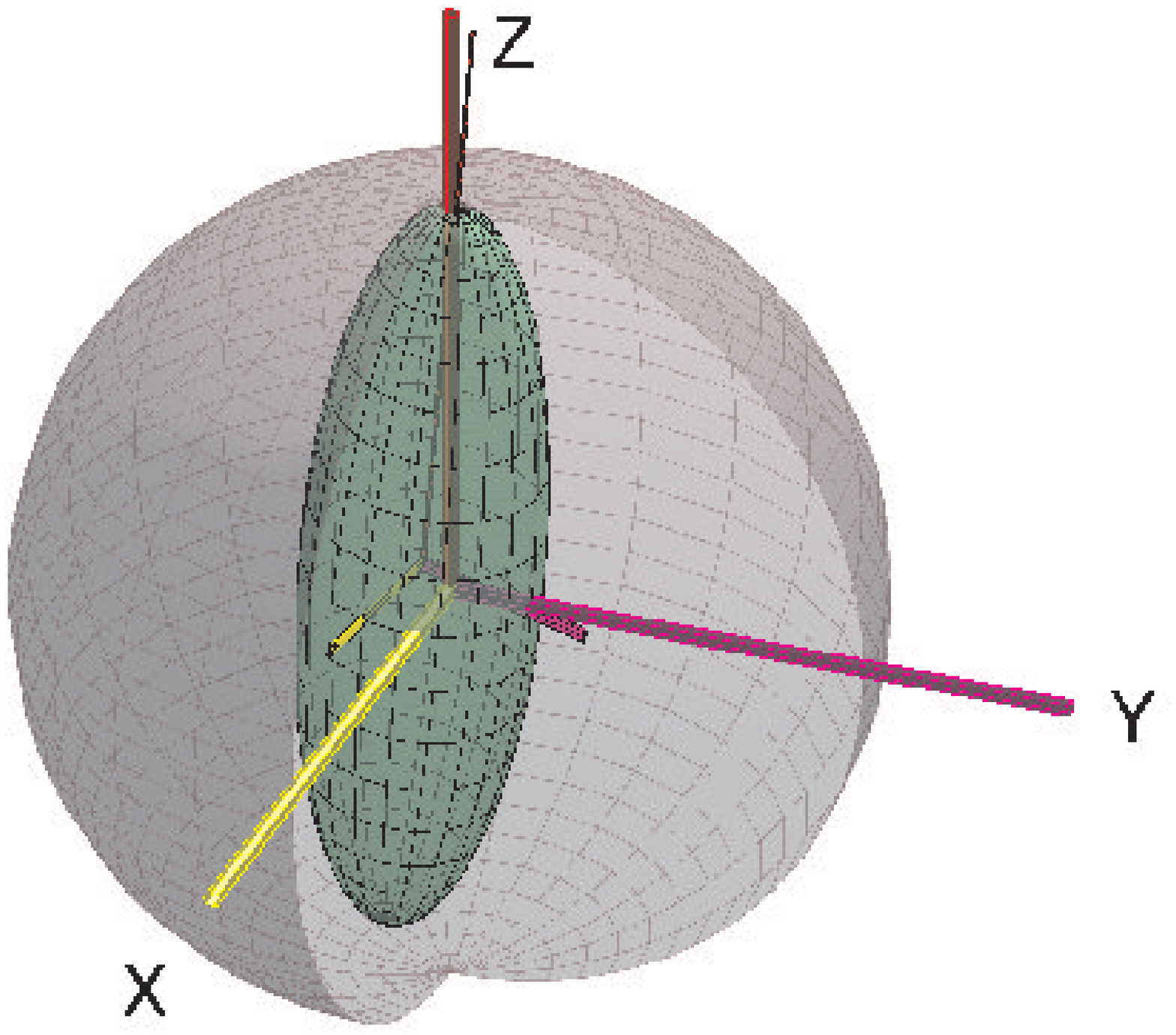,height=33mm}}{{(b)}}
\caption{80ns results:(a) Bloch sphere representation of process from experimental data. (b) Bloch sphere representation of process obtained
from $\tilde{\chi}$. }
\end{center}
\label{fig3}
\end{figure}

\end{document}